# Gate defined quantum dot realized in a single crystalline InSb nanosheet

Jianhong Xue,[1,3] Yuanjie Chen,[1] Dong Pan[2] Ji-Yin Wang,[1] Jianhua Zhao,[2] Shaoyun Huang,[1,*] and H. Q. Xu[1,4,*]

[1]*Beijing Key Laboratory of Quantum Devices, Key Laboratory for the Physics and Chemistry of Nanodevices and Department of Electronics, Peking University, Beijing 100871, China*

[2]*State Key Laboratory of Superlattices and Microstructures, Institute of Semiconductors, Chinese Academy of Sciences, P.O. Box 912, Beijing 100083, China*

[3]*Academy for Advanced Interdisciplinary Studies, Peking University, Beijing 100871, China*

[4]*Division of Solid State Physics, Lund University, Box 118, S-221 00 Lund, Sweden*

[*]Corresponding authors. Emails: hqxu@pku.edu.cn; syhuang@pku.edu.cn

## ABSTRACT

Single crystalline InSb nanosheet is an emerging planar semiconductor material with potential applications in electronics, infrared optoelectronics, spintronics and topological quantum computing. Here we report on realization of a quantum dot device from a single crystalline InSb nanosheet grown by molecular-beam epitaxy. The device is fabricated from the nanosheet on a Si/SiO$_2$ substrate and the quantum dot confinement is achieved by top gate technique. Transport measurements show a series of Coulomb diamonds, demonstrating that the quantum dot is well defined and highly tunable. Tunable, gate-defined, planar InSb quantum dots offer a renewed platform for developing semiconductor-based quantum computation technology.

With their small bandgap, high electron mobility, small electron effective mass, large electron Landé g-factor, and strong spin-orbit interaction, III−V semiconductor InSb nanostructures have attracted an intensive interest. In the past few years, the investigation of InSb nanostructures has focused on epitaxially grown InSb nanowires[1-17], and various InSb nanowire devices, such as field-effect transistors[2-5], single[6-8] and double quantum dot devices[9,10], and semiconductor–superconductor hybrid quantum devices[11-17], have been realized and been widely anticipated to have potential applications in spintronics, semiconductor-based quantum technology, and in detection and manipulation of Majorana Fermions for topological quantum computing. Recently, free-standing, single-crystalline InSb nanosheets[18,19] have been successfully grown by molecular-beam epitaxy (MBE) and have been found to exhibit superior quantum transport properties when compared to other layered semiconductor materials. Due to the two-dimensional nature, electrical and quantum devices fabricated from these InSb nanosheets offer a renewed platform for developing novel high-speed nanoelectronic, infrared optoelectronic, spintronic, and quantum computing technologies.

In this letter, we report on realization of a quantum dot device from an MBE-grown, free-standing, single crystalline InSb nanosheet. The nanosheet is of high quality and has a high electron mobility. The device is fabricated from the nanosheet on a Si/SiO$_2$ substrate using standard semiconductor fabrication techniques and the quantum dot confinement is achieved by top gate technique. The fabricated device is characterized by transport measurements in a He$^3$/He$^4$ dilution refrigerator at a base temperature of 20 mK. The measured charge stability diagram shows a series of Coulomb diamonds, demonstrating that the shape of quantum dot and the number of electrons in the dot can be tuned efficiently with the top and Si back gates.

The device studied in this work is fabricated from a free-standing, high-quality single-crystalline pure-phase InSb nanosheet. InSb nanosheets are grown by MBE on top of InAs nanowires on a Si(111) substrate. The process starts by depositing a thin layer of Ag on the Si substrate in the MBE chamber and then annealing is carried out *in situ* to generate Ag nanoparticles. Using these Ag nanoparticles as seeds, InAs nanowires are then grown[20-23]. InSb nanosheets are grown on top of these InAs nanowires by abruptly switching the group-V source from As to Sb and with an increased Sb flux. The grown InSb nanosheets can be up to several micrometers in size and down to ~10 nm in thickness. See Ref. 18 for further details about the growth process and structural properties of our MBE grown InSb nanosheet materials.

For device fabrication, the MBE grown InSb nanosheets are mechanically transferred onto a highly doped Si substrate covered with a 300-nm-thick layer of SiO$_2$ on top. The Si substrate and the SiO$_2$ layer are later used as a global back gate and gate dielectric to the nanosheet, respectively. After transferring, standard electron-beam lithography (EBL) is used for pattern definition of electrical contacts on a selected InSb nanosheet. It is determined after completing all the measurements that the selected nanosheet has a width of $W$~430 nm, a length of $L$~1160 nm and a thickness of $t$~30 nm. Metal contacts are then fabricated by deposition of 5-nm-thick Ti and 90-nm-thick Au in an electron-beam evaporator (EBE) and lift-off process. Here we note that before loading the sample into the EBE chamber, the sample is immersed in a H$_2$O-diluted (NH$_4$)$_2$S$_x$ solution to remove the surface contamination and native surface oxide, and the thin layer of Ti used here is to promote metal adhesion to the nanosheet and the substrate. After another step of EBL, a 20 nm thick HfO$_2$ layer is deposited on top of the nanosheet by atomic layer deposition (ALD) and lift-off. Finally, top gates are fabricated by a third step of EBL followed by EBE of 5 nm Ti and 90 nm Au and lift-off process. Figure 1 (a) shows a schematic view of the device structure and the inset of Figure 1(b) shows a scanning electron microscope (SEM) image of the fabricated InSb nanosheet device, where the four top gates, labeled by G1, G2, G3 and G4, and the source and drain contacts labeled by S and D, are clearly displayed.

The fabricated device is measured in a He$^3$/He$^4$ dilution refrigerator. Figure 1(b) shows measured source-drain current $I_{ds}$ as a function of back gate voltage $V_{bg}$ at temperature $T$= 20 mK and source-drain bias voltage $V_{ds}$ = 0.5 mV with all top gates being grounded (i.e., by setting all voltages applied to the top gates at $V_{G1}$=$V_{G2}$=$V_{G3}$=$V_{G4}$=0). It is seen that the device is a typical n-type transistor—current $I_{ds}$ is pinched off at $V_{bg}$=−5.5 V and shows a saturation at positive back gate voltages. The electron mobility of the InSb nanosheet can be extracted from the measured $I_{ds}$-$V_{bg}$ curve using equation, $\mu = \frac{L}{W \times C_{bg}} \frac{1}{V_{ds}} \times \frac{dI_{ds}}{dV_{bg}}$, where $C_{bg}$ is the unit area capacitance, which can be estimated from $C_{bg} = \frac{\varepsilon_0 \varepsilon}{d}$ with $\varepsilon_0$ being the vacuum permittivity, $\varepsilon = 3.9$ the dielectric constant of SiO$_2$, and $d$ = 300 nm the thickness of SiO$_2$. The extracted mobility from the measured $I_{ds}$-$V_{bg}$ curve is ~11000 cm$^2 \cdot$V$^{-1} \cdot$s$^{-1}$, demonstrating that the InSb nanosheet device exhibits an excellent high-speed electronic performance.

The performances of the top gates measured at $V_{bg}$=−2 V. Hereafter, we set $V_{bg}$=−2 V, as indicated by a red star in Figure 1(b), in order to get a sufficient carrier density in the nanosheet, and thus any influences of disorders and contact resistances can be reduced when we define a quantum dot in the nanosheet[24,25]. Figure 2(a) shows the measured source-drain current $I_{ds}$ as a function of gate voltage $V_{G1}$ at $V_{ds}$=0.1 mV with all other top gates being grounded, i.e., $V_{G2}$=0

V, $V_{G3}$=0 V, and $V_{G4}$=0 V. Here with decreasing $V_{G1}$, a quick drop of $I_{ds}$ to a low level is observed, which means that the carriers in the InSb nanosheet beneath gate G1 can be electrostatically depleted. Here, a complete depletion of carriers across the InSb nanosheet can not be achieved using gate G1 only. The residual $I_{ds}$ is the current passing through the InSb nanosheet area not covered by gate G1. Figure 2(b) shows the measured $I_{ds}$ as a function of $V_{G2}$ at $V_{ds}$=0.1 mV and a fixed $V_{G1}$=−3.6 V, indicated by a red star in Figure 2(a), with the other two top gates being grounded as $V_{G3}$=0 V and $V_{G4}$=0 V. Here, with decreasing $V_{G2}$, $I_{ds}$ drops to zero, which means the channel current in the InSb nanosheet can be electrostatically cut off using gates G1 and G2. Figure 2(c) shows the measured $I_{ds}$ as a function of $V_{G4}$ at $V_{ds}$=0.1 mV and $V_{G1}$=−3.6 V with gates G2 and G3 being grounded. Here, similar as seen in Figure 2(b), the channel current can also be cut off using gates G1 and G4. Figure 2(d) shows the measured $I_{ds}$ as a function of $V_{G3}$ at $V_{ds}$=0.1 mV and $V_{G1}$=−3.6 V with gates $G_2$ and G4 being grounded. Here, it is seen that the channel current in the InAs nanosheet can not be completely cut off using gates G1 and G3. This is as expected, because the distance between the two gates is ~180 nm, which is too large to form a potential barrier across the InSb nanosheet.

Based on the measured gate performances shown in Figures 1(b) and 2, a quantum dot can be created in the InSb nanosheet by setting voltages applied to the back and top gates to proper values. In this work, we choose to define a quantum dot in the nanosheet by setting $V_{bg}$= −2 V, $V_{G1}$=−3.6 V, $V_{G2}$=−5.3 V, and $V_{G4}$=−4.7 V, as indicated by red stars in Figures 1(b), 2(a), 2(b) and (c). Gate G3 is used to tune the dot shape and the number of electrons in the dot. Figure 3(a) shows the source-drain $I_{ds}$ measured for such defined quantum dot as a function of $V_{G3}$ at $V_{ds}$=0.1 mV. The measured $I_{ds}$ are characterized by densely distributed sharp current peaks which are typical for electron transport through a quantum dot in the Coulomb blockade regime[2,26-28]. It is seen that the current peaks distribute differently in the low, negative gate voltage region and the high, positive gate voltage region—the current peak spacing ($\Delta V_{G3}$) is much larger in the low, negative gate voltage region than in the high, positive gate voltage region [cf. the zoom-in plot shown in the inset of Figure 3(a) for the current peak spacing in a high, positive gate voltage region]. Thus, the gate capacitance ($C_{G3}$) to the dot is different in the two gate voltage regions.

Figure 3(b) shows the differential conductance measured for the quantum dot as a function of $V_{ds}$ and $V_{G3}$ (charge stability diagram) over a gate voltage $V_{G3}$ region from −1.5 to 0.4 V. Here, well-defined Coulomb diamond structures, characteristics for a quantum dot in the Coulomb blockade regime, are observed. However, the addition energies extracted from the charge stability diagram are very different in the low, negative gate voltage region and high, positive gate voltage region. For a detailed analysis, we show, in Figures 3(c) and 3(d), zoom-in plots of the charge stability diagram in a low, negative gate voltage region marked by black dashed rectangle I and in a high, positive gate voltage region marked by black dashed rectangle II. In both Figures 3(c) and 3(d), five Coulomb diamonds are shown. The addition energies $E_{add}$ extracted from the Coulomb diamonds shown in Figure 3(c) are marked by red dots in Figure 3(e), giving an average addition energy of $\bar{E}_{add}^{I}$~3.1 meV. The addition energies $E_{add}$ extracted from the Coulomb diamonds shown in Figure 3(d) are marked by blue dots in Figure 3(f), giving an average addition energy of $\bar{E}_{add}^{II}$~1.5 meV. It is seen that the addition energies in the low, negative gate voltage region is about twice larger than that in the high, positive gate voltage region, implying that the dot size is significantly smaller in region I than region II.

To have a better understanding of how the quantum dot is formed in the InSb nanosheet in our device, we have simulated the electron potential profile in the nanosheet using commercial available program COMSOL[29,30]. In the simulation, the same layer structure as in our experiment is assumed and the boundary condition is chosen to set the voltages in the regions of the InSb nanosheet that are contacted by the source and drain electrodes to zero. In the experiment, a current flow is observed when the gates are grounded, indicating the existence of a finite density of electrons in the nanosheet at zero gate voltages. To take this into account in the simulation, the experimental threshold voltages of the back gate ($V_{bg}^{th} = -5.4$ V), top gate G1 ($V_{G1}^{th} = -2.1$ V), top gate G2 ($V_{G2}^{th} = -3.1$ V), and top gate G4 ($V_{G4}^{th} = -3.0$ V), at which the electrons start to be depleted, are extracted from the measurements shown in Figures 1(b), 2(a), 2(b), and 2(c).[31] The gate voltages input into the simulation, $V_{bg}^{s} = 3.4$ V, $V_{G1}^{s} = -1.5$ V, $V_{G2}^{s} = -2.2$ V, and $V_{G4}^{s} = -1.7$ V, are the values after calibrations against the corresponding experimental threshold voltages. Here, we note that because in the measurements, the voltage applied to top gate G3 is an experimental variable, we make no effort to extract its experimental threshold voltage, but rather input reasonable gate voltage values of gate G3 directly into the simulations. Figures 4(a) and 4(b) show the simulated electron potential profiles in the InSb nanoshheet at two voltages applied to top gate G3. In Figure 3(a) where the simulation is carried out by setting $V_{G3}^{s} = -0.4$ V, it is seen that the electron potential in the InSb nanosheet region beneath gate G3 is high and positive, and no electrons could be populated there. Thus, a quantum dot of approximately circular shape located in the middle region of the nanosheet is formed. This is the situation corresponding to the gate voltage region I in the measurements shown in Figure 3. In Figure 4(b) where the simulation is carried out by setting $V_{G3}^s = 0.36$ V, it is seen that the electron potential in the InSb nanosheet region beneath gate G3 is low and negative, and the electrons in the InSb nanosheet region beneath gate G3 are not completely depleted. Thus, a quantum dot of spoon-like shape is formed. This is the situation corresponding to the gate voltage region II in the measurements shown in Figure 3. The above results of simulations show that when tuning G3 gate voltage around $V_{G3}^s = -0.4$ V ($V_{G3}^s = 0.36$ V), Coulomb diamonds with a larger (smaller) addition energy would be observed in the measured stability diagram, in consistence with the measurements shown in Figure 3. Our simulations also demonstrate that in gate voltage region I, gate G3 is situated on a side of the quantum dot, while in region II, gate G3 covers a part of the quantum dot. Thus, the gate capacitance ($C_{G3}$) to the dot is smaller in gate voltage region I than in gate voltage region II.

In conclusion, a quantum dot device is realized in an emerging single crystalline, semiconductor InSb nanosheet. The nanosheet is grown by MBE and is of high crystalline quality. The quantum dot is fabricated on a Si/SiO$_2$ substrate and is defined by gating technique. The measured charge stability of the fabricated quantum dot device exhibits a series of well-defined Coulomb diamonds over a large range of voltages applied to a control gate. The extracted addition energies from the measured Coulomb diamonds show that the quantum dot is very different in size in the low, negative control-gate voltage region and the high, positive control-gate voltage region. The electron confinement potential in the InSb nanosheet is simulated for the fabricated device. This difference in quantum dot size has been confirmed by the simulations and the overall dot shape has been visualized. Our work demonstrates that highly tunable quantum dot system can be built from InSb monolayers for potential applications in semiconductor quantum dot based quantum computation technology developments.

This work is supported by the Ministry of Science and Technology of China through the National Key Research and Development Program of China (Grant Nos. 2017YFA0303304, 2016YFA0300601, 2017YFA0204901, and 2016YFA0300802), and the National Natural Science Foundation of China (Grant Nos. 11874071, 91221202, 91421303, 61504133, and 11274021). HQX also acknowledges financial support from the Swedish Research Council (VR).

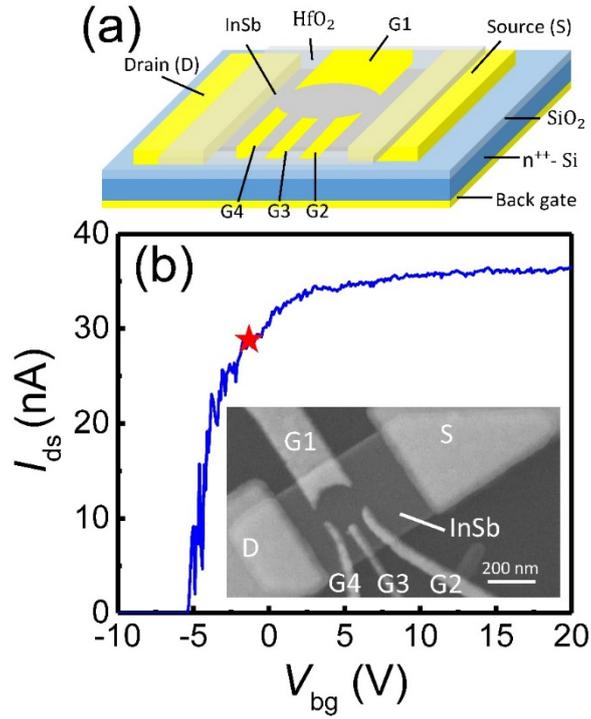

Figure 1. (a) Schematic view of the device structure. (b) SEM image of the device measured in this work (inset) and source-drain current $I_{ds}$ measured for the device as a function of back gate voltage $V_{bg}$ (transfer characteristics) at temperature $T$=20 mK and source-drain bias voltage $V_{ds}$=0.5 mV (main figure). In the device structure schematic and the SEM image of the measured device, the source and drain contacts are labeled as S and D, and the four top gates are labeled as G1, G2, G3 and G4. In the measurements for the transfer characteristics, the four top gates are grounded, namely, all the top gate voltages are set at $V_{G1}$=0 V, $V_{G2}$=0 V, $V_{G3}$=0 V, and $V_{G4}$=0 V.

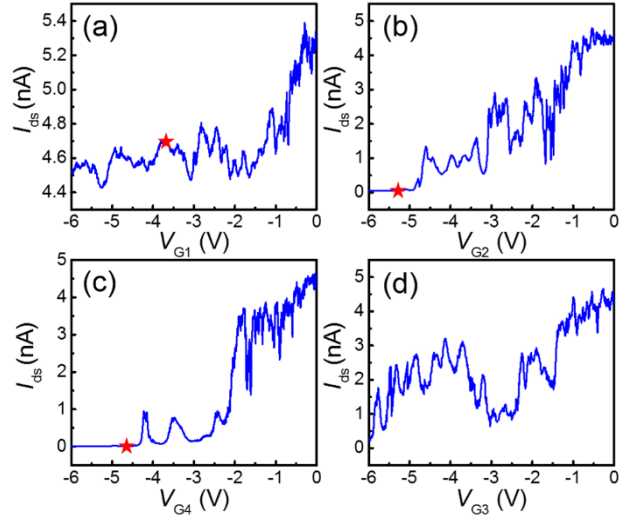

Figure 2. Transfer characteristics of the top gates. (a) Source-drain current $I_{ds}$ measured for the device as a function of $V_{G1}$ at $T=20$ mK and $V_{ds}=0.1$ mV with other gate voltages set at $V_{bg}=-2$ V, $V_{G2}=0$ V, $V_{G3}=0$ V, $V_{G4}=0$ V. (b) The same as in (a) but for $I_{ds}$ as a function of $V_{G2}$ with $V_{G1}=-3.6$ V, $V_{G3}=0$ V, and $V_{G4}=0$ V. (c) The same as in (a) but for $I_{ds}$ as a function of $V_{G4}$ with $V_{G1}=-3.6$ V, $V_{G2}=0$ V, and $V_{G3}=0$ V. (d) The same as in (a) but for $I_{ds}$ as a function of $V_{G3}$ with $V_{G1}=-3.6$ V, $V_{G2}=0$ V, and $V_{G4}=0$ V.

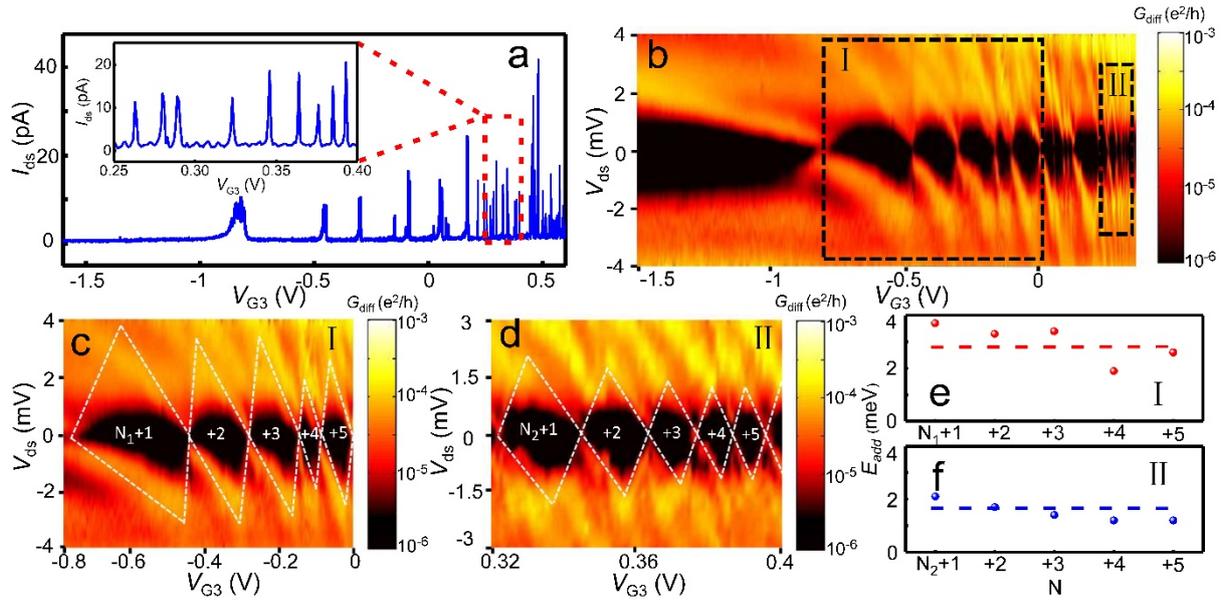

Figure 3. (a) Coulomb current oscillations of the quantum dot defined by setting $V_{bg}=-2$ V, $V_{G1}=-3.6$ V, $V_{G2}=-5.3$ V, and $V_{G4}=-4.7$ V measured against $V_{G3}$ at $T=20$ mK and $V_{ds}=0.1$ mV. The inset shows a zoom-in plot in a high, positive $V_{G3}$ region. (b) Differential conductance $G_{diff}$ measured for the device as a function of $V_{ds}$ and $V_{G3}$ (charge stability diagram) at $T=20$ mK. (c) Zoom-in plot of the measured charge stability diagram in the region marked by black rectangle I in (b). (d) Zoom-in plot of the measured charge stability diagram in the region marked by black rectangle II in (b). (e) Addition energies extracted from the measurements shown in (c), i.e., in region I. The red dashed line indicates the mean value of the addition energies. (f) Addition energies extracted from the measurements shown in (d), i.e., in region II. The blue dashed line indicates the mean value of the addition energies.

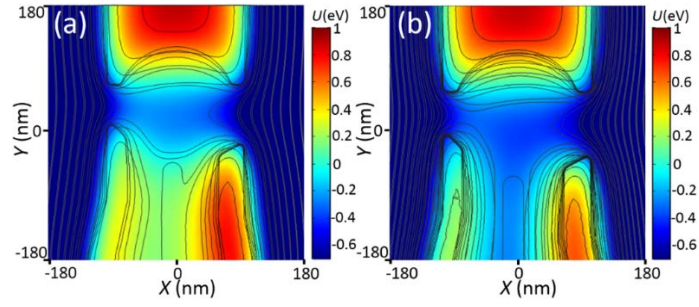

Figure 4. Simulated electron potential landscapes in the InSb nanosheet region surrounded by the top gates. (a) Simulated potential landscape at $V_{bg}^s = 3.5$ V, $V_{G1}^s = -1.5$ V, $V_{G2}^s = -2.2$ V, $V_{G4}^s = -1.7$ V and $V_{G3}^s = -0.4$ V. (b) Simulated potential landscape at $V_{bg}^s = 3.5$ V, $V_{G1}^s = -1.5$ V, $V_{G2}^s = -2.2$ V, $V_{G4}^s = -1.7$ V and $V_{G3}^s = 0.36$ V.